\documentclass[prd,aps,superscriptaddress,floatfix,nofootinbib,eqsecnum,twocolumn]{revtex4-1}
\pdfoutput=1

\usepackage{amsfonts}
\usepackage{amsmath}
\usepackage{amssymb}
\usepackage{bm}
\usepackage{dcolumn}
\usepackage{graphicx}   
\usepackage[latin1]{inputenc}
\usepackage{latexsym}
\usepackage{rotating}
\usepackage{hyperref}
\usepackage{subfigure}
\usepackage{color}
\usepackage{changes}

\begin{document}

\title{Gravitational particle production in loop quantum cosmology}

\author{L. L. Graef}
\email{leilagraef@if.uff.br}
\affiliation{Instituto de F\'{\i}sica, Universidade Federal Fluminense,
24210-346 Niter\'oi, RJ, Brazil}

\author{Rudnei O. Ramos}
\email{rudnei@uerj.br}
\affiliation{Departamento de F\'{\i}sica Te\'orica, Universidade do
Estado do Rio de Janeiro, 20550-013 Rio de Janeiro, RJ, Brazil}

\author{G.~S.~Vicente}\email{gustavo@fat.uerj.br}
\affiliation{Faculdade de Tecnologia, Universidade do Estado do Rio de
  Janeiro, 27537-000 Resende, RJ, Brazil}

\begin{abstract}

We investigate the gravitational particle production in the bounce
phase of  loop quantum cosmology (LQC). We perform both analytical and
numerical analysis of the particle production  process in a LQC
scenario with a Bunch-Davies vacuum initial condition in the contracting
phase. We obtain that  if we  extend the validity of the dressed
metric approach beyond the limit of small backreaction in which it is
well justified, this process would lead  to a radiation dominated
phase in the  preinflationary phase of LQC. Our results indicate that
the test field approximation, which is required in the truncation
scheme used in the dressed metric approach, might not be a valid
assumption in a LQC scenario with such initial conditions. 
 
\end{abstract}

\maketitle 

\section{Introduction} 

It is well known that in the description of the very early Universe
and tracing back the time evolution,  we enter in a Planck scale
regime where quantum gravity effects can dominate.    On such scales,
modifications of general relativity (GR) are expected to be important.
In the theoretical explorations of the early Universe, one generally
uses the  {}Friedmann-Lema\^{\i}tre-Robertson-Walker (FLRW) solutions
to the Einstein  equations (with appropriate matter sources) as
background spacetimes.  However, the FLRW spacetimes of interest are
incomplete in the past due to  the big bang singularity, when matter
fields and spacetime curvature diverge.  General relativity is not a
suitable theory once curvature reaches the Planck scale, where quantum
effects become important.  Therefore, we cannot expect reliable
results for quantum fields evolving in  classical backgrounds. The
natural way is to look for a quantum theory of gravity, which, e.g.,
can  provide the quantization of spacetime.

There are some viable candidate models of quantum gravity that can be
able to describe the physics of the very early Universe.  Some of
these models share the feature of avoiding the initial singularity,
which results in a nonsingular  bouncing Universe. 
 Although inflation is the dominant paradigm for the early Universe,
there are alternative ideas~\cite{Lilley:2015ksa,Creminelli:2010ba,Brandenberger:2009jq}, like several bouncing models, which can agree with  current cosmological observations as well as inflation does.
In these models,
the quantum effects are important at the Planck scale and  are
responsible for the bounce, but become less important as the Universe
evolves away from Planck scale, thus eventually  recovering GR. Among
these quantum gravity candidates, loop quantum gravity (LQG) provides
a promising avenue in this
direction~\cite{Bojowald:2001xe,Ashtekar:2006rx,Ashtekar:2006uz,Ashtekar:2006wn,Ashtekar:2007em,Singh:2009mz,Ashtekar:2006es,Szulc:2006ep}.
loop quantum cosmology (LQC) arises as the result of applying the
principles of LQG to cosmological
settings~\cite{Ashtekar:2011ni,Barrau:2013ula,Agullo:2016tjh}.  In LQC
the quantum geometry creates a repulsive effective force that is
totally negligible at  low spacetime curvature, but grows rapidly in
the Planck regime, overwhelming the classical  gravitational
attraction. In cosmological models, while Einstein's equations hold to
an excellent degree of approximation  at low curvature, they receive
important corrections in the Planck regime. As a consequence of these
corrections, any time a curvature invariant grows to the Planck scale,
quantum geometry effects dilute it, resolving the problems of
singularities of GR.

In addition to the bounce and contraction phases, the inflationary
scenario admits an extension for LQC in the expanding phase.
 Inflation can be important to put several models in agreement with observations in the context of LQC (see, for example, Ref.~\cite{Zhu:2017jew}).
 Several  authors  have  investigated  the  naturalness of an inflationary phase after the bounce~\cite{Ashtekar:2011rm,Linsefors:2013cd,Martineau:2017sti,Bolliet:2017czc}. Despite  the  different viewpoints, in
LQC by starting with a scalar field with a  typical potential
considered in inflationary models, after the bounce an inflationary
phase  will  almost inevitably sets in.  In this
context, it is important to analyze the physical implications of
preinflationary dynamics in LQC, i.e., the quantum evolution from the
bounce  to the onset of the slow roll.  

An important aspect to investigate in bouncing cosmologies is the
Parker mechanism of gravitational  particle production
(GPP)~\cite{Parker:1968mv,Parker:1969au}, which has shown to be very
efficient in the bounce phase of several
models~\cite{Quintin:2014oea,Tavakoli:2014mra,Haro:2015zda,Hipolito-Ricaldi:2016kqq,Celani:2016cwm,Scardua:2018omf,Zago:2018huk}.
If we assume that the spacetime is the Minkowski one both at early
(prebounce) and at late (postbounce) times, but underwent a period
of expansion during an intermediate time interval, then the initial
Minkowski vacuum state of a test scalar field $\chi$ will evolve
nontrivially during the intermediate time interval to a final state
which is not equal to the final time Minkowski vacuum. {}From the
point of view of the final Minkowski frame, the final state contains
$\chi$ particles. The same phenomenon applies to linear cosmological
perturbations that evolve in a similar way to test scalar fields on
the cosmological background.  It was shown in
Ref.~\cite{Quintin:2014oea} that in the matter bounce scenario the
process of GPP is sufficient to produce a hot early Universe. 
 The matter bounce scenario was also investigated in the context of LQC in Ref.~\cite{WilsonEwing:2012pu}.
It was
also shown in Ref.~\cite{Hipolito-Ricaldi:2016kqq} that GPP, under some conditions, can also be responsible for the
emergence of a hot thermal state in the Universe in the new ekpyrotic
model, thus avoiding the need  to introduce an additional reheating
phase.  In Refs.~\cite{Celani:2016cwm,Scardua:2018omf,an,DM}, the GPP
was also analyzed in the context of  other bouncing
cosmologies~\footnote{In Refs.~\cite{julio,JE} this production was
  analyzed specifically in the context of LQC, but in a different
  framework than the one considered here.}.  

In this work, we compute the GPP via the Parker
mechanism~\cite{Parker:1968mv,Parker:1969au} in the preinflationary
phase of  LQC.   We compute the energy density stored in the produced
particles and compare it with the background energy density, which we
consider to be dominated by the kinetic energy of the inflaton field
before  the slow-roll  phase.  In our framework, we will consider
gravity coupled to  scalar fields and study the dynamics of quantum
fields  on quantum cosmological spacetimes.

Since the interest relies on the dynamics of quantized fields
propagating on these quantum cosmological backgrounds, one needs a
quantum gravity extension of the standard field equations. The
extension we are going to consider lies in the framework of the
dressed metric approach, which allows for a description of the  field
mode equations in a form analogous to the classical
one~\cite{outroAgullo2013}. The dressed metric approach considers a
truncation scheme that is assumed to be well justified in the cases in
which the backreaction of the produced modes are negligible. By
computing the backreaction of the gravitationally produced particles
in the preinflationary phase of LQC, we will be able to test the
validity of the dressed metric approach in this scenario. 

This paper is organized as follows.  In Sec.~\ref{sec2}, we
describe the background dynamics of the LQC model considered here. In
Sec.~\ref{sec3}, we analyze the gravitational particle production
in LQC. In Sec.~\ref{sec4}, we present the  analytical and the
numerical results for  the energy density of the particles
produced. {}Finally, our concluding remarks are presented in
Sec.~\ref{sec5}.

\section{Background Model}
\label{sec2}
 
We consider LQC as our cosmological  scenario  from which the
Einstein's equations receive explicit modifications in the Planck
regime. The spatial geometry in LQC is encoded in the volume of a
fixed fiducial cubic cell, rather than the scale factor $a$, and is
given by 
\begin{equation}
v = \frac{{\cal V}_0 a^{3} m_{\rm Pl}^2}{2\pi\gamma},
\label{eqv}
\end{equation}
where ${\cal V}_0$ is the comoving volume of the fiducial cell,
$\gamma$ is the Barbero-Immirzi parameter~\cite{Meissner:2004ju} of
LQC, whose numerical value is given by $\gamma\simeq 0.2375$, $m_{\rm
  Pl}\equiv 1/\sqrt{G} = 1.22 \times 10^{19}\,$GeV is the Planck mass
and $G$ is the Newton constant of gravitation. The conjugate momentum
to $v$ is denoted by $b$ and it is given by $b=-4\pi\gamma P_{(a)}/(3
a^2 {\cal V}_0 m_{\rm Pl}^2)$,  where $P_{(a)}$ is the conjugate
momentum to the scale factor.  

The solution of the LQC effective equations implies that the Hubble
parameter $H$ can be written as
\begin{equation}
H=\frac{1}{2 \gamma \lambda} \sin(2 \lambda b),
\label{eqH}
\end{equation}
where $\lambda =(48\pi^2 \gamma^2/m_{\rm Pl}^4)^{1/4}$ and $b$ ranges
over $(0, \pi/\lambda)$.   The energy density, $\rho$, relates to the
LQC variable $b$ through $\rho = 3 m_{\rm Pl}^2 \sin^2(\lambda
b)/(8\pi\gamma^2 \lambda^2)$.  Then, the {}Friedman equation in LQC
assumes the form~\cite{Ashtekar:2011rm}
\begin{equation}
\frac{1}{9}\left(\frac{\dot{v}}{v}\right)^{2} \equiv H^2 =
\frac{8\pi}{3 m_{\rm Pl}^2} \rho \left(1- \frac{\rho}{\rho_{\rm c}}
\right),
\label{Hubble}
\end{equation}
where $\rho$ is the energy density,  $\rho_c = 3 m_{\rm
  Pl}^2/(8\pi\gamma^2 \lambda^2)\approx 0.41m_\mathrm{Pl}^4$ is the
critical density in LQC and the dots denote derivatives with respect
to the cosmic time. {}For $\rho \ll \rho_{\rm c}$, we recover GR as
expected. The expression~(\ref{Hubble}) holds independently of the
particular characteristics of the inflationary regime. We can see that
the singularity is avoided in this model and,  when the energy density
approaches the critical density, the Universe undergoes a bounce with
$H=0$.

We consider a cosmological scenario in which the energy content of the
Universe is dominated by the inflaton field, with a potential
$V(\phi)$. We also consider in our scenario an extra scalar field
$\chi$ with no interactions to standard model (SM) fields and also no
direct coupling to the inflaton. Hence, the field $\chi$ only couples
to $\phi$ and to the SM particles gravitationally. 
 There are many well motivated candidates for spectator fields in cosmology.  {}For example, dark matter can well be completely
decoupled  from the visible matter (and also from the inflaton) and
interact only gravitationally~\cite{Peebles:1999fz}.  
The production of
spectators can all be gravitational (typically, at the end of
inflation in the usual scenarios), or they can also be produced
through graviton mediated scatterings (in which case, their
interactions with other fields are naturally suppressed by Planck mass
factors).  Another example of a spectator field in cosmology is the
curvaton (see, e.g., Ref.~\cite{Lyth:2001nq}), which is essentially, in its simplest
implementations, decoupled from the inflaton and not contributing to
the energy density during inflation, but it can become important later
on.  We believe that spectator fields are well motivated in cosmology
and the GPP of them in the present study is of interest.
Here we also extend the hypothesis of $\chi$ being a spectator field to
in the prebounce and around the bounce phases. Thus, $\chi$ has negligible contribution to
the background dynamics during these phases.  However, due to the
marked change in the metric evolution at around the bounce, we expect
that $\chi$ will be produced gravitationally and it can have important
effects in the postbounce subsequent evolution.  We want to determine
how important will be this contribution of the GPP of $\chi$ particles
to the subsequent postbounce preinflationary dynamics.

Due to the quantum nature of $H$, we can explicitly observe the
evolution of the scale factor during the cosmological phase close to
the bounce, where the Universe is dominated by the inflaton field,
which is described by  a barotropic fluid with equation of state
$p=\omega \rho$.  This solution reads (see, e.g.,
Ref.~\cite{Wilson-Ewing:2013bla}):
\begin{eqnarray}
a(t)=a_B \left[1+\frac{6\pi\rho_c}{m_\mathrm{Pl}^2}(1+\omega)^2
  t^2\right]^{\frac{1}{3(1+\omega)}}.
\end{eqnarray}

In this work, we will consider the case of a bounce dominated by the
inflaton kinetic energy, which behaves as stiff matter, i.e., like a
fluid with equation of state $\omega\approx 1$. During this phase, the
scale factor is given by
\begin{equation}\label{astiff}
a=a_{B}\left(1+\gamma_{B}\frac{t^{2}}{t_\mathrm{Pl}^{2}}\right)^{1/6},
\end{equation}
where $\gamma_{B}\equiv 24\pi\rho_{c}/m_\mathrm{Pl}^{4}\simeq 30.9$
and $t_\mathrm{Pl}\equiv 1/m_\mathrm{Pl}$ is the Planck time. This can
be accomplished with the homogeneous scalar field inflaton $\phi$
subjected to a potential $V(\phi)$, whose evolution equation reads
\begin{equation}\label{eqphi}
\ddot \phi + 3 H \dot \phi + V_{,\phi}=0.
\end{equation}

In this scenario, with the bounce dominated by the kinetic energy of
$\phi$, the evolution of the Universe after the contraction, and prior
to preheating, can be divided in three different phases according to
the behavior of the equation of state of the dominant fluid (see,
e.g., Ref.~\cite{Zhu:2017jew}): The bouncing phase, the transition
phase and the slow-roll inflation phase. In the bounce phase,  which
lasts from  the bounce until  around $t\sim  10^{4} t_{\rm Pl}$,   the
evolution of $a(t)$ is independent of the inflationary model, since
the inflaton potential energy density is negligible around the bounce.
In the transition phase, the kinetic energy of the inflaton decreases
very rapidly  (about 12 orders from its initial Planck scale), and the
equation of state changes suddenly from $w(\phi)\sim 1$ to
$w(\phi)\sim -1$. The transition phase is very short compared to the
other phases. It starts at  $t\sim  10^{4}\, t_{\rm Pl}$ and last
until $t\sim  10^{5}\, t_{\rm Pl}$. After that, the slow-roll phase
starts with the inflaton potential energy dominating. {}From the
bounce time to the beginning of inflation (i.e., during the
preinflationary phase) the Universe expands around $4-5$
\textit{e}-foldings~\cite{Zhu:2017jew,Graef:2018ulg}. As we will see bellow,
the dynamics of the  fields equations is different in each of these
three phases.

As already mentioned above, the aim of this work is to see how the GPP
of an spectator scalar field $\chi$ can affect the  preinflationary
phase of LQC. At the same time,  this will also provide us with means
to test the validity of the dressed metric approach in this
scenario. We begin by describing the mechanism of particle production
in the following section.

\section{Particle Production in Curved Space-times}
\label{sec3}

We consider the standard procedure  of describing quantum fields on
classical, though curved, spacetimes. However, this strategy cannot be
directly justified in the quantum gravity era, where curvature and
matter densities are of Planck scale. Nevertheless, using techniques
from LQG, the standard theory was extended providing us with means to
overcome this limitation~\cite{outroAgullo2013}. 

\subsection{Dressed metric approach}

In LQG, we do not  have  the analog of the full Einstein's equations
to perturb. Several strategies have been developed to overcome this
issue. Here, we follow the mainstream strategy in LQG, which consists
in first truncating the classical theory in a manner appropriate to
the physical problem under consideration. Then, the quantization is
carried out considering the  underlying quantum geometry of
LQG. {}Finally, the consequences of the resulting framework are
derived.  The full phase space is truncated, keeping only the FLRW
background with first-order inhomogeneous perturbations.  The
truncated phase space is described by $\Gamma_{\rm
  Trun}=\Gamma_{0}\times \tilde{\Gamma}_{1}$,      where $\Gamma_{0}$
is the four-dimensional phase space for the FLRW background and
$\tilde{\Gamma}_{1}$ is the phase space of gauge invariant
perturbations~\cite{Agullo2012, Agullo2013, AKL2009}. {}For a Universe
dominated by the scalar field $\phi$, the background phase space
$\Gamma_{0}$ is coordinatized by [$\nu,b;\phi,p_{(\phi)}$] and carries
a single  Hamiltonian constraint which implies in the equation,
\begin{equation}\label{S0}
S_{0}[N_{\rm hom}]=N_{\rm hom}\left[-\frac{3b^{2}\nu}{4\gamma} +
  \frac{p^{2}_{(\phi)}m_{\rm Pl}^{2}}{4 \pi \gamma \nu}+\frac{2 \pi
    \gamma \nu}{m_{\rm Pl}^{2}}V(\phi)\right]=0,
\end{equation}
where $N_{\rm hom}$ represents an homogeneous lapse.  By considering
in the above equation the choice $N_{\rm hom}=1$, it gives the
evolution in cosmic time, while $N_{\rm hom}=a$ gives it in conformal
time, $N_{\rm hom}=a^{3}$ in harmonic time and $N_{\rm hom}=a^{3}{\cal
  V}_{0}/p_{(\phi)}$ corresponds to using the inflaton field as ``time
clock" (see, e.g., Ref.~\cite{Agullo2012} for more details).
 
Since the phase space of the truncated system is a product, the
Hilbert space of quantum states has the form
$\mathcal{H}=\mathcal{H}_{0}\otimes \mathcal{H}_{1}$.  The Hilbert
space of background fields, $\mathcal{H}_{0}$, consists of wave
functions $\Psi_{0}(\nu,\phi)$. The evolution of this quantum state is
governed by the LQC quantum Hamiltonian constraint.  Thus, the wave
function $\Psi_{0}(\nu, \phi)$  is subjected to the constraint
$\hat{S}_{0}\Psi_{0}=0$ from the Dirac quantization procedure. {}From
this constraint, we are led to the equation $-i\partial_{\phi}
\Psi_{0}(\nu,\phi)= \hat{H}_{0} \Psi_{0}(\nu,\phi)$~\cite{Agullo2012},
where $\hat{H}_{0}$ is a self-adjoint operator whose explicit form
will not be needed here. 

Therefore, the evolution of this quantum state with respect to the
emergent time variable $\phi$ is governed by the LQC quantum
Hamiltonian constraint. Among several states $\Psi_{0}(a,\phi)$	 in
the LQC Hilbert space, one is interested in a state that is sharply
peaked around a classical trajectory at late times. Evolving this
state by using the LQC quantum Hamiltonian constraint, it has been
shown that it remains sharply peaked during the whole dynamical
trajectory, even deep in the  Planck
era~\cite{Agullo2012}. Consequently, the evolution of the peak of such
states can be described by an effective trajectory that is governed by
the effective equations.  Now  quantum perturbations  propagate on
quantum geometries which are  regular and free of singularities, being
the energy density  bounded above by the critical value $\rho_{c}
\simeq 0.41m_\mathrm{Pl}^4$. 

On the other hand,  the classical dynamics on the full $\Gamma_{\rm
  Trun}$ is not generated by a constraint, therefore, one cannot
recover the quantum dynamics for the total system by imposing a
quantum constraint. We do this by first proceeding in the homogeneous
sector as mentioned, by  reinterpreting the quantum Hamiltonian
constraint  as an evolution equation  in the homogeneous sector, and 
then, we ``lift'' the resulting quantum trajectory to the full Hilbert
space $\mathcal{H}$ (see Ref.~\cite{Agullo2012}) considering the
truncated space. Nevertheless, in order for the truncation scheme to
be valid, we need to ensure that the energy density stored in the
perturbations is much smaller than the energy density of the
background~\cite{Agullo2013}, i.e.,
$\rho_\mathrm{pert}/\rho_\mathrm{bg}\ll1$,  all the way back to the
bounce. Only then  we would be assured of a self-consistent solution,
justifying our truncation, which ignores the backreaction. Otherwise,
one would have to perform a full quantum gravity theory. 

In the scenario considered here, in addition to the dominant inflaton
field, we consider a spectator scalar field $\chi$, which has absent,
or negligible interactions to the other components of the Universe.
We follow the analysis of the gravitational  production of this scalar
field $\chi$ in the framework of the dressed metric
approach~\cite{Agullo2012,Agullo2013,AKL2009,outroAgullo2013}, which
has the advantage of allowing the description of the main equations in
a form analogous to the classical ones.  In the dressed metric
approach, the equation of motion of the operators representing scalar
perturbations and the scalar field equations are formally the same as
the usual equations appearing in a classical spacetime in GR. 
We analyze the GPP associated with this scalar field $\chi$. Considering
that $\chi$ has a sufficiently small or negligible mass (compared,
e.g., to $H$ in the postbounce phase), thus, $\chi$ behaves
essentially as radiation during the whole preinflationary phase.

In the test field approximation, the dynamics of  fields propagating
on the quantum geometry, which is described by the quantum state
$\Psi_{0}$, behave as  propagating in  a quantum modified effective
geometry described by the following dressed metric~\cite{Agullo2012,
  AKL2009},
\begin{equation}
\tilde{g}_{ab} dx^{a}dx^{b}=\tilde{a}^2(-d\tilde{\eta}^2 +dx_{i}
dx^{i}),    
\end{equation}
where the dressed scale factor $\tilde a$ and the dressed conformal
time $\tilde \eta$ are given, respectively, by
\begin{equation}
\tilde{a} = \left(\frac{ \langle \hat{H}^{-1/2}_{0} \hat{a}^{4}
  \hat{H}^{-1/2}_{0} \rangle } {\langle \hat{H}^{-1}_{0} \rangle
}\right)^{1/4},
\end{equation}
and
\begin{equation}
d\tilde{\eta} = \langle \hat{H}^{-1/2}_{0}\rangle  \langle
\hat{H}^{-1/2}_{0} \hat{a}^{4}  \hat{H}^{-1/2}_{0} \rangle^{1/2} d\phi
.
\end{equation}
In the latter equations, $\hat{H}_{0}$ is the background Hamiltonian
and the expectation values are taken with respect to the background
quantum geometry state, which is given by $\Psi_{0}(a,\phi)$.  In this
approach, the equations of motion of the operators representing scalar
and tensor perturbations are formally the same as the equations
appearing in classical spacetimes, which in {}Fourier space read,
\begin{equation}\label{mutilde}
\mu_{k}''(\tilde{\eta}) + \left[k^{2} - \frac{\tilde{a}''}{\tilde{a}}
  + \tilde{U}(\tilde{\eta})\right]\mu_{k}(\tilde{\eta})=0,
\end{equation}
where primes here denotes derivative with respect to the conformal
time, $\mu_k(\tilde{\eta}) = z\,R_k$, with $R_k$ denoting the comoving
curvature perturbation, $z(\tilde{\eta})= a\dot{\phi}/H$ and
\begin{equation}
\tilde{U}(\tilde{\eta})=\frac{\langle\hat{H}^{-1/2}_{0} \hat{a}^{2}
  \hat{U}(\phi) \hat{a}^{2}  \hat{H}^{-1/2}_{0}\rangle}{\langle
  \hat{H}^{-1/2}_{0} \hat{a}^{4} \hat{H}^{-1/2}_{0}\rangle}.
\end{equation}
The quantities $\tilde{a}$, $\tilde{\eta}$ and
$\tilde{U}(\tilde{\eta})$  represent the quantum expectation values in
the background state $\Psi_{0}(a,\phi)$. However, for sharply peaked
background states, the dressed effective quantities are well
approximated by their peaked values $a$, $\eta$ and $U(\eta)$. Then,
the equation of motion for scalar modes, Eq.~\eqref{mutilde}, becomes
\begin{equation}\label{mueom}
\mu_{k}''({\eta}) + \left[k^{2} - \frac{{a}''}{{a}} +
  {U}({\eta})\right]\mu_{k}({\eta})=0,
\end{equation}
where $U=a^{2}(\it{f}^{2}V(\phi) + 2\it{f}V_{,\phi}(\phi)+V_{,\phi
  \phi}(\phi))$ and $\it{f} \equiv \sqrt{2\pi
  G}\dot{\phi}/\sqrt{\rho}$. The effective potential $U$ is negligible
in the bounce and transition phases and it can then be
neglected~\cite{Zhu:2017jew}. Thus, the equation of motion in the
bounce and transition phases can be simply written as
\begin{eqnarray}\label{muk}
\mu_k''(\eta) + \left[k^2 -
  \frac{a''(\eta)}{a(\eta)}\right]\mu_k(\eta)=0.
\end{eqnarray}

 Despite its use in the present study, it is important to mention 
that there are other alternatives to the dressed metric approach. Other important approaches for treating perturbations in LQC are, e.g., the hybrid quantization approach (see, for example, Refs.~\cite{Gomar:2014faa,hybrid} and references therein), and the deformed algebra approach (see, for example, Ref.~\cite{Schander:2015eja} and references therein).

\subsection{Gravitational particle production}

By considering the evolution of the {}Fourier modes $\mathcal{X}_k$ of
the scalar field $\chi$, in the dressed metric approach, the form of
its equation of motion is quite analogous to the classical spacetime
equations.  Also, by neglecting the $\chi$ mass, the evolution
equation for the conformal rescaled field mode,  $\chi_k\equiv a
\mathcal{X}_k$, simply takes the same form as that for the
perturbation modes Eq.~(\ref{muk}).  Hence, we can simply identify
$\mu_k \to \chi_k$ in Eq.~(\ref{muk}), giving 
\begin{eqnarray}\label{Phik}
\chi_k''(\eta) + \left[k^2-
  \frac{a''(\eta)}{a(\eta)}\right]\chi_k(\eta)=0.
\end{eqnarray}
Equation~\eqref{Phik} represents a set of uncoupled oscillators with
time variable frequency $\sqrt{k^2  - a''/a}$.  The time dependence of
$a''/a$ stems from the evolution of the scale factor $a(\eta)$, which
parametrizes the  background evolution. Therefore, for each instant
$\eta$ we define a different vacuum.
Parker~\cite{Parker:1968mv,Parker:1969au} established conditions for
the definition of a time dependent particle number operator $n(\eta)$.
It can be defined if its vacuum expectation value varies slowly as
possible with time as the expansion rate of the Universe is
sufficiently slow.  Also, this expansion period must occur between two
''Minkowskian'' vacuum states.  One evaluates the effect of GPP
occurring between these vacuum states. The mathematical treatment is
summarized below.  

The Hamiltonian for $\chi_k(\eta)$ in terms of its {}Fourier modes
reads,
\begin{eqnarray}\label{H}
\!\!\!\!\!\!\!\!\!H(\eta) \!=\! \int d^3k  \left(  2 E_k
\hat{a}_{\vec{k}}^\dagger \hat{a}_{\vec{k}} + F_{\vec{k}}
\hat{a}_{\vec{k}} \hat{a}_{-\vec{k}} + F_{\vec{k}}^*
\hat{a}_{\vec{k}}^\dagger \hat{a}_{-\vec{k}}^\dagger \right),
\end{eqnarray}
where
\begin{eqnarray}
&&E_k(\eta)= \frac{1}{2}|\chi_k'(\eta)|^2 +
  \frac{\omega_k^2}{2}|\chi_k(\eta)|^2,
\label{Ew}
\\ &&F_k(\eta)=\frac{1}{2}(\chi_k'(\eta))^2+ \frac{\omega_k^2}{2}
(\chi_k(\eta))^2,
\label{Fw} 
\end{eqnarray}
where $\omega_k=k$ in the massless approximation for the $\chi$ field.
We diagonalize the Hamiltonian performing the following Bogoliubov
transformation:
\begin{eqnarray}
\hat{b}_{\vec{k}}=\alpha_k(\eta)\hat{a}_{\vec{k}} +\beta_k^*(\eta)
    {\hat{a}_{-\vec{k}}}^\dagger,
\end{eqnarray}
where the Bogoliubov coefficients $\alpha_k(\eta)$ and $\beta_k(\eta)$
satisfy the constraint  $ |\alpha_k(\eta)|^2  - |\beta_k(\eta)|^2=1$
due to normalization of the modes.  The resulting diagonal Hamiltonian
reads:
\begin{eqnarray}\label{Hdiag}
H(\eta)= \int d^3k \  \omega_k\  b_{\vec{k}}^\dagger\  b_{\vec{k}},
\end{eqnarray}
and Eq.~(\ref{Ew}), for instance, becomes
\begin{eqnarray}\label{Ek}
E_k(\eta)=\omega_k\left[ \frac{1}{2} + |\beta_k(\eta)|^2\right].
\end{eqnarray}  
Defining the vacuum states $|0_{(a)}\rangle$ and $|0_{(b)}\rangle$
such that $a_{\vec{k}}|0_{(a)}\rangle=b_{\vec{k}}|0_{(b)}\rangle=0$,
we can compute the expectation value of the number operator
$\hat{N}_{\vec{k}}^{(b)}=b_{\vec{k}}^\dagger\  b_{\vec{k}}$ in the
vacuum $|0_{(a)}\rangle$:
\begin{eqnarray}\label{N}
n_k(\eta)=
\langle_{(a)}0|\hat{N}_{\vec{k}}^{(b)}|0_{(a)}\rangle=|\beta_k(\eta)|^2.
\end{eqnarray}
We observe that $|\beta_k(\eta)|^2$ is exactly the particle number per
mode.  Therefore, from Eq.~\eqref{Ek}, we obtain that
$E_k(\eta)=\omega_k\left[1/2+n_k(\eta)\right]$ is the energy
contribution of GPP.  The Bogoliubov coefficients relate the initial
Minkowskian vacuum states $\chi_k^{(i)}$ to the final ones
$\chi_k^{(f)}$ in the following way:
\begin{eqnarray}\label{modesbogoliubov}
\chi_k^{(f)}(\eta)=\alpha_k \chi_k^{(i)}(\eta) + \beta_k
\chi_k^{(i)*}(\eta).
\end{eqnarray}
When $\beta_k=0$, there is no particle production and the constraint
$|\alpha_k(\eta)|^2  - |\beta_k(\eta)|^2=1$ gives  $\alpha_k=1$ and,
then, $\chi_k^{(i)}=\chi_k^{(f)}$ for the whole evolution.

{}From the definition of the particle number per mode, Eq.~(\ref{N}),
we can obtain the total particle number density  $n_p(\eta)$ and the
total energy density $\rho_p(\eta)$ produced.  The total particle
number density can be defined as the limit of
$\sum_{k}|\beta_k(\eta)|^2$ in a box of side $L\to\infty$ divided by
the volume  $a^3(\eta)L^3$, which gives
\begin{eqnarray}\label{nv0}
n_p(\eta)&=&\frac{1}{a^3(\eta)
  L^3}\left(\frac{L}{2\pi}\right)^3\int\limits_0^\infty
d^3k\   n_k(\eta)  \nonumber
\\ &=&\frac{1}{2\pi^2a^3(\eta)}\int\limits_0^\infty dk \ k^2
|\beta_k(\eta)|^2.
\end{eqnarray}
Likewise, the total energy density $\rho_p(\eta)$ can by defined as
the momentum sum of $\omega_k \, n_k (\eta)$, which results in
\begin{eqnarray}\label{rhov0}
\rho_p(\eta)=\frac{1}{2\pi^2a^4(\eta)}\int\limits_0^\infty dk\,
k^2\,\omega_k  |\beta_k(\eta)|^2.
\end{eqnarray} 
Equation~(\ref{rhov0}) allows us to obtain the energy density
associated with the particles that are gravitationally produced in the
Universe. 

It is important to mention that the energy density of produced
particles can also be obtained from the expectation value of the
energy-momentum tensor of $\chi$ field at any time $\eta$, which
reads~\cite{Bunch:1980vc}
\begin{eqnarray}\label{rhoEM}
\rho_p^{\mathrm{EM}}(\eta)&=&\frac{1}{4\pi^2a^4(\eta)}\int\limits_0^\infty
dk\ k^2 \left\{\frac{}{}|\chi_k'(\eta)|^2 + \omega_k^2|\chi_k(\eta)|^2
\right.  \nonumber \\ &+& \left.
\left(\frac{a'}{a}\right)^2|\chi_k(\eta)|^2-\frac{a'}{a}
\left[\chi_k^*(\eta)\chi_k(\eta)\right]'\right\}.
\end{eqnarray} 
{}Far from the bounce, when the expansion rate is negligible, the
energy difference due to produced particles with respect to the
initial vacuum reduces to Eq.~\eqref{rhov0}. 

In the following we will directly solve the equations of motion for
the  fields in each cosmological phase. 

\subsection{The scalar field equation of motion}

In order to perform a preliminary analytical analysis, we will neglect
the backreaction effect of the produced particles on the cosmological
background and we will solve the equation of motion for the field
modes $\chi_k$, Eq.~(\ref{Phik}), for each cosmological
phase. Backreaction effects will be considered explicitly when we
numerically solve for the modes later in
Sec.~\ref{numerical_backreaction}.

To obtain the solution for the  modes  $\chi_k$, we need to solve
Eq.~(\ref{Phik}) in each phase (bounce, transition and slow-roll
inflationary phase). We will not compute the particle production in
the previous matter contraction phase since, independently of the
details of the contracting phase, particles are effectively produced
only after they get rid of the influence of the potential $a''/a$,
which happens soon after the bounce phase in the LQC model here
considered.  The behavior of the modes in each phase  depends on which
term, $k^{2}$ or $a''/a$, dominates the time dependent  frequency in
Eq.~\eqref{Phik}. To analyze the relevant modes, it is useful to
define the characteristic length $\lambda=\sqrt{a/a''}$, which plays a
role analogous to that of the comoving Hubble radius. We also define
the characteristic momentum $k_B=\sqrt{a''/a}|_{t=t_B}$, which is the
physical energy at the bounce.  Note that from Eq.~(\ref{astiff}), we
have that 
\begin{equation}
k_B \equiv \sqrt{\gamma_B/3}\, m_{\rm Pl} \simeq 3.21\, m_{\rm Pl},
\label{kB}
\end{equation}
where we have set $a_B =1$ at the bounce.  The relevant field modes
$\chi_k$ are those with $k\approx k_{B}$, as these are the modes
expected to give the main contribution to the GPP. However, the
dynamics of these modes have different behavior when they are inside
and outside the characteristic length $\lambda$. We will focus on the
modes  that begin well inside $\lambda$ in the contracting phase, exit
$\lambda$ during the bounce phase and then enter $\lambda$ again in
the transition phase and then on.  We consider the initial
Bunch-Davies (BD) vacuum state in the contracting phase. In the
following subsections we describe the solution for $\chi_{k}(\eta)$ in
each cosmological phase.

\subsubsection{Bounce phase}

Since the equations for the momentum modes for the perturbations,
Eq.~(\ref{muk}), and that for the modes $\chi_k$, Eq.~(\ref{Phik}),
are essentially identical, we can quite conveniently adapt the results
already derived by the authors in Ref.~\cite{Zhu:2017jew} and used in
there to solve for Eq.~(\ref{muk}) to get the bounce contribution to
the power spectrum. This starts by realizing that Eq.~(\ref{Phik}) can
be seen as analogous to a type of the Schr\"{o}dinger equation and
where the term $a''(\eta)/a(\eta)$ acts like a potential, which
behaves as an effective barrier during the bouncing phase. This
potential can be well approximated by a P\"{o}schl-Teller potential at
the bounce, i.e., $a''(\eta)/a(\eta) \approx k_{B}^{2}\,
\mathrm{sech}^{2}[\sqrt{6}\,k_{B}(\eta-\eta_{B})]$.  In this case, the
solution of Eq.~\eqref{Phik} can be put in the form of the solution of a standard hypergeometric equation and it is given by~\cite{Zhu:2017jew}
\begin{eqnarray}
\!\!\!\!\!\!\!\!\!\!\!\!\!\! \chi_{k}(\eta) &=& a_{k} x^{ik/(2\sqrt{6}
   k_{B})}(1-x)^{-ik/(2\sqrt{6} k_{B})}   \nonumber \\ &\times&
 _{2}F_{1} (a_{1} - a_{3} +1, a_{2}-a_{3}+1, 2 -a_{3}, x)  \nonumber
 \\ &+& \; b_{k} [x(1-x)]^{-ik/(2\sqrt{6} k_{B})} \;  _{2}F_{1}(a_{1},
 a_{2}, a_{3}, x),
\label{bouncesol}
\end{eqnarray}
where $x\equiv x(\eta)=\{1+\exp[-2\sqrt{6}\, k_{B}\, (\eta-\eta_{B})]\}^{-1}$,
\begin{eqnarray}
&&a_{1} \equiv \frac{1}{2}\left(1+\frac{1}{\sqrt{3}}\right) -
  \frac{ik}{\sqrt{6}k_{B}},  \\ &&a_{2} \equiv
  \frac{1}{2}\left(1-\frac{1}{\sqrt{3}}\right) -
  \frac{ik}{\sqrt{6}k_{B}},  \\ &&a_{3} \equiv 1 -
  \frac{ik}{\sqrt{6}k_{B}},
\end{eqnarray}
and $a_{k}$ and $b_{k}$ are integration constants determined by the
initial conditions. These will be explicitly determined below.

\subsubsection{Transition phase}

In the transition phase, on the other hand,  the term $k^2$ dominates
over $a''/a$ in Eq.~\eqref{Phik}. The solution of the equation of
motion in this phase is given by
\begin{equation}\label{transitionsol}
\chi_{k}= \frac{1}{\sqrt{2k}}(\tilde{\alpha}_{k} e^{-ik\eta} +
\tilde{\beta_{k}} e^{ik\eta}),
\end{equation}
where $\tilde{\alpha_{k}}$ and $\tilde{\beta_{k}}$ are two more
integration constants and also determined when matching the different
solutions.

\subsubsection{Slow-roll inflationary phase}

After the transition phase, in the slow-roll inflationary phase, the
Universe is no longer in the quantum regime and the equation of motion
reduces to the usual relativistic one,
\begin{equation}\label{eom2}
\chi_{k}''(\eta) + \left(k^{2} - \frac{\nu^{2}+1/4}{\eta^{2}}
\right)\chi_{k}(\eta)=0,
\end{equation}
where 
\begin{equation}
    \nu^{2}=\eta^{2}\frac{a''(\eta)}{a(\eta)}+\frac{1}{4}.
\end{equation}
In this phase the approximate analytical  solution of Eq.~\eqref{eom2}
is given in terms of the Hankel functions as
\begin{equation}\label{inflsol}
\chi_{k}(\eta) \approx \frac{\sqrt{-\pi \eta}}{2}[\alpha_{k}
  H_{\nu}^{(1)}(-k\eta) + \beta_{k} H_{\nu}^{(2)}(-k\eta)],
\end{equation}
where ${\alpha_{k}}$ and ${\beta_{k}}$ are again two other integration
constants.

\subsubsection{Matching phases}\label{matching}

To determine the coefficients $\alpha_{k}$ and $\beta_{k}$ in
Eq.~(\ref{inflsol}), one needs to match the solutions of each phase in
their intermediate regions. During the contracting phase, close to the
bounce, all the relevant  modes are well inside the characteristic
length $\lambda$. Then, we can choose the BD vacuum state as the
initial conditions of the modes. Therefore,
\begin{equation}
    \chi_{k}^{in}(\eta) \approx \frac{1}{\sqrt{2k}}e^{-ik\eta}.
\end{equation}
We can consider the solution of the bounce phase,
Eq.~\eqref{bouncesol}, in the limit $\eta - \eta_{B} \gg 0$.  By
comparing the resulting equation with the above initial condition, we
obtain that the coefficients $a_k$ and $b_k$ in Eq.~\eqref{bouncesol}
are given by
\begin{equation}
a_{k}=0, \; \; \;\; \; \;
b_{k}=e^{ik\eta_{B}}/\sqrt{2k}.
\end{equation}
Next, by comparing the solution for $\chi_k(\eta)$ in the bounce
phase, Eq.~\eqref{bouncesol}, with the solution in the transition
phase, Eq.~\eqref{transitionsol}, we can obtain the expressions for
$\tilde{\alpha}_{k}$ and $\tilde{\beta}_{k}$. These, compared to the
solution in the slow-roll, Eq.~\eqref{inflsol}, in the limit $-k\eta
\rightarrow \infty$, give, after some some algebra, 
that~\cite{Zhu:2017jew}
\begin{eqnarray}
&& \alpha_{k}= \tilde{\alpha}_{k}=\frac{\Gamma(a_{3})
    \Gamma(a_{1}+a_{2}-a_{3})}{\Gamma(a_{1})\Gamma(a_{2})} e^{2 ik
    \eta_{B}}, \\ &&\beta_{k}= \tilde{\beta}_{k}=\frac{\Gamma(a_{3})
    \Gamma(a_{3}-a_{1}-a_{2})}{\Gamma(a_{3}-a_{1})\Gamma(a_{3}-a_{2})}.
\end{eqnarray}
{}From the previous results, it then follows that 
\begin{equation}\label{beta}
|\beta_{k}|^{2} =
\frac{1}{2}\left[1+\cos\left(\frac{\pi}{\sqrt{3}}\right)\right] {\rm
  csch}^{2}\left(\frac{\pi k}{\sqrt{6}k_{B}}\right).
\end{equation}
The coefficients of the slow-roll phase (far future) and the ones of
the contracting phase (far past) are related through
Eq.~\eqref{modesbogoliubov}. This relation gives that the coefficient
$\beta_{k}$ of Eq.~\eqref{beta} is exactly the Bogoliubov
coefficient. As we have already discussed, the quantity
$|\beta_{k}|^{2}$ is the number of particles produced per mode $k$,
$n_{k}$. By computing this quantity, we can infer the quantity of
particles produced since the initial BD vacuum in the beginning of the
contracting phase~\footnote{The presence of particles produced before
  the beginning of inflation have consequences to the power spectrum
  of the model, which receives a correction factor such that it is
  written as $\Delta_\mathcal{R}(k)= |\alpha_{k} +\beta_{k}|^{2}
  \Delta_\mathcal{R}^{GR}(k)$, where $\Delta_\mathcal{R}^{GR}$ is the
  GR form of the spectrum~\cite{Agullo2013,Zhu:2017jew}.   A fast
  oscillatory behavior of the power spectra arises from the final
  interference term, due to the rapidly changing relative phase. This
  oscillation is so fast in $k$ that in any realistic observations
  they would be averaged out~\cite{Agullo2013}.  Consequently, the LQC
  power spectrum can be simply given by rescaling the standard one
  from general relativity by a factor of $(1+2|\beta_{k}|^{2})$. The
  same equations of motion described in this section were also solved
  in Refs.~\cite{Zhu:2017jew, Agullo2013} in the context of the
  curvature perturbation. }. This will be computed in the next section
both analytically, when neglecting the backreaction effect of the
produced particles on the background evolution, and numerically, by
fully including its effect.

\section{Results}
\label{sec4}

We next compute particle production due to the gravitational effects
both analytically and numerically.  {}From the analytical point of
view, we make use of the results of the previous
Sec.~\ref{matching}. In this case, the produced particles take no part
in the dynamics; i.e., there are no backreaction effects.  On the
other hand, in the numerical setting, we consistently compute the
particle production and include its effects in the Universe dynamics,
such that the backreaction effects are self-consistently taken into
account.

\subsection{Analytical results}
\label{secIVA}

The energy density of relativistic $\chi$ particles that are produced
due to the bounce is determined by Eq.~(\ref{rhov0})  with
Eq.~\eqref{beta}.  When performing the momentum integral in this
equation, we must establish both ultraviolet (UV) and infrared (IR)
cutoffs. This point requires some discussion since there is not a
consensus on the choice of these integration limits. {}For instance,
in  Ref.~\cite{Agullo2013}, the integration limits were chosen such
that the range of frequencies integrated are defined by the window of
observable modes in the cosmic microwave background (CMB). This choice
implies in a $k_{\rm min}$ much higher than $k_{B}$. As discussed in
Ref.~\cite{Agullo2013}, the dynamics of modes for such high values of
$k$ is largely insensitive to the background geometry, so they
essentially  evolve as if they were in flat spacetime. In this case
the evolved state is indistinguishable from the BD vacuum at the onset
of inflation. In addition,  there is nothing, in principle, to prevent
particle production already starting at the scale $\lambda$ of the
effective horizon, right after the bounce. The particles that are
produced at that moment should contribute to the energy density,
specially if they are relativistic (radiationlike). In fact, if a too
low IR cutoff $k_{\rm min}$ is taken such to be much higher than
$k_{B}$, we lose essentially all particle production, since $k_B$ is
exactly the scale around which (and below it)  we expected that
most of the gravitational particle production, due to the bounce,
effectively happens.  The associated modes may not be observed today,
but they should contribute to the energy density during all the
preinflationary era, where their energy density is not redshifted
away too fast. {}For this reason, here we choose a different approach.
{}For our purpose of estimating the energy density stored in the
particles produced, we will be  interested in the modes such that
$k_\mathrm{min}(\eta) \leq k\leq k_{B}$, where
$k_\mathrm{min}(\eta)=\lambda^{-1} \equiv \sqrt{a''(\eta)/a(\eta)}$
for all $\eta$.  The produced modes are effectively considered as
particles after they reenter the horizon.  Those are the modes which
exits and reenter the effective horizon $\lambda$ during the
preinflationary phase, and then they will only reexit $\lambda$
later, in the slow-roll inflationary phase. Hence, those are the modes
which  are expected to give the largest contribution to the particle
production, since those are the ones that exit  $\lambda$ and
reenters it before inflation starts. Our UV cutoff will then be given
by $k_{B}$, while $\lambda^{-1}$ yields the natural IR cutoff in the
present problem.  Therefore, the energy density of the particles
created is given by Eq.~\eqref{rhov0},  which with the appropriate IR
and UV cutoffs, can be written as
\begin{eqnarray}
\label{rhofinal}
\rho_p(\eta)=\frac{1}{2\pi^2a^4(\eta)}\int_{k_\mathrm{min}(\eta)}^{k_B}
dk\, k^2\, n_k(\eta)\, \omega_k,
\end{eqnarray} 
which upon using $n_k \equiv |\beta_k|^2$, with $|\beta_k|^2$ given by
Eq.~\eqref{beta} and $\omega_k\sim k$ for relativistic $\chi$
particles, we obtain that
\begin{equation}\label{intfromzero}
\rho_{p}(\eta) = \frac{1+\cos\left(\frac{\pi}{\sqrt{3}}\right) }{4
  \pi^{2} a^{4}(\eta)} \int_{k_\mathrm{min}(\eta)}^{k_B} dk\, k^{3}\,
    {\rm csch}^{2}\left(\frac{\pi k}{\sqrt{6}k_{B}}\right).
\end{equation}
Since the integral is dominated by the UV limit $k_B$, we can simply
set the lower limit of integration to zero  in the above
equation. Thus, Eq.~(\ref{intfromzero}) gives 
\begin{equation}\label{rhoGPP}
\rho_{p}(\eta) \simeq 4.5\times 10^{-3} \frac{k_{B}^{4}}{a^{4}(\eta)}
\simeq 0.012 \frac{ m_{\rm Pl}^4}{a^{4}(\eta)},
\end{equation}
where we have used the result~\eqref{kB} for
$k_B$. Equation~(\ref{rhoGPP}) gives the energy density of particles
produced after the main modes reenter $\lambda$ after the bounce,
i.e., when $a(\eta)>1$.  As already previously mentioned, the relevant
modes for GPP of relativistic $\chi$ particles are the ones such that
$k<k_{B}$, which are the ones that exit and reenter $\lambda$ during
the bounce phase. These modes will only reexit $\lambda$ again in the
slow-roll inflationary phase. Therefore, we can neglect the particle
production after the end of the bounce phase until the beginning of
the inflationary phase, since the relevant modes for GPP are all
inside $\lambda$ during this period.

We are considering the cases where the kinetic energy density of the
inflaton is the dominant energy component at the bounce.   Thus, the
background energy evolves like stiff matter, $\rho_{\rm back} =
\rho_c/a^6$.  On the other hand, the $\chi$ GPP evolves like
relativistic matter $\sim 1/a^4$, as seen in Eq.~(\ref{intfromzero}).
Let us first estimate an upper bound on the energy density of GPP by
imposing that it remains subdominant up to the beginning of
inflation. We write the energy density of particles produced as
$\rho_{p}= \rho_{\chi}\, a^{-4}$, where $\rho_{\chi}$ is the energy
density of particles produced soon after the bounce (when the main
modes reenter $\lambda$). The elapsed \textit{e}-foldings between the  bounce
and the start of inflation in LQC is around $N_{\rm preinfl}= 4 - 5$
\textit{e}-folds (see, e.g., Ref.~\cite{Zhu:2017jew}). Thus, by requiring that
at the start of inflation $\rho_{\chi} <  \rho_\mathrm{bg}$, we
readily find the condition  $\rho_\chi \lesssim 2 \times 10^{-5}m_{\rm
  Pl}^{4}$ for the energy density of relativistic GPP produced around
the bounce {\it not to dominate} the energy content of the Universe up
until before inflation.  Let us now estimate $\rho_\chi$ from the
result given by Eq.~(\ref{rhoGPP}). We can take as an estimate  the
maximum of GPP to happen around the time $t_s$ where $a''/a=0$ in the
postbounce phase,  which can be estimated from Eq.~(\ref{astiff}) to
be $t_s \simeq 0.3 t_{\rm Pl}$. Thus, from Eq.~(\ref{rhoGPP}) we can
estimate that $\rho_\chi \equiv \rho_p(t_s) \simeq 5 \times 10^{-3}
m_{\rm Pl}^{4}$.  This result is about 2 orders of magnitude larger
than the upper bound estimated above for the relativistic GPP not to
dominate until before the beginning of inflation.  Given the previous
discussion about the validity of the truncation scheme used in the
dressed metric approach, we see that what the above result indicates
is that this standard procedure cannot be justified in this scenario,
being necessary a full quantization approach.

However, the above analytical estimate for the energy density of GPP
is expected to overestimate the particle production, since the
backreaction of this GPP is fully absent  in the derivation of
$\beta_k(\eta)$ that was presented in the previous section and, when
included, should affect the evolution dynamics of the background,
potentially decreasing the net GPP.   In the next section we compute
the particle production by including continuously the backreaction of
the GPP on the cosmological background.  We then verify whether we can
extend the validity of the dressed metric approach beyond the limit
which it is strictly justified.

\subsection{Numerical results}
\label{numerical_backreaction}

The inclusion of the backreaction effects of the particles that are
continuously produced on the cosmological background can only be done
numerically.  As in any problem of GPP, we have to properly account
for how the initial conditions are set, in particular, for the  field
modes.  We can consider two different sets of initial conditions for
the modes. One is the BD vacuum, imposed at the contracting phase
right before the quantum bounce, and the other is the fourth-order
adiabatic vacuum state\cite{Bunch:1980vc}  set at the bounce.  In
Ref.~\cite{Zhu:2017jew}, it was shown that, in the scenarios
considered here, they are essentially equivalent and lead to the same
results. In the following we take advantage of this finding and we are
going to consider the initial BD vacuum state  in our numerical
analysis, which makes the numerics much simpler.

As we have mentioned in the latter subsection, we will be  interested
in the modes such that $k_\mathrm{min}(\eta) \leq k\leq k_{B}$, where
$k_\mathrm{min}(\eta) \equiv\sqrt{a''(\eta)/a(\eta)}$. Therefore,
using these UV and IR limits, the energy density of produced
particles, reads like Eq.~(\ref{rhofinal}) from the previous
subsection.  The expression for the number density $n_k$ is given
through $E_k$, Eq.~(\ref{Ew}), expressed in terms of the field modes
$\chi_k$ and their time derivatives.  The field modes $\chi_k(\eta)$
are explicitly obtained by numerically solving their equation of
motion, given by Eq.~\eqref{Phik}.  If we fix the scale factor as the
one given by Eq.~\eqref{astiff} and compute the GPP, this procedure
must give results similar to Eq.~\eqref{intfromzero}.  This is the
case where there is no backreaction in the Universe dynamics. However,
we are interested in including the effect of particle production on
its dynamics, which demands us to add to the numerical system the
equation for the stiff matter field evolution, Eq.~\eqref{eqphi}, and
to include the energy density of produced particles in the {}Friedmann
equation, Eq.~\eqref{Hubble}. 

It is important to draw attention to some of the aspects of the GPP
considered here.  We compute particle production soon after the
bounce.  We consider as particles  the modes which get inside
$\lambda$, after the squeezing during the bounce, i.e.,  modes that
satisfy $k^2<a''/a$ during the bounce phase for some time interval.
Then, the energy contribution from particle production is computed at
the instant these modes reenter $\lambda$.  {}For each instant, we
consider the modes which have entered $\lambda$ and integrate on these
modes. Near the bounce, modes with $k\lesssim k_B$ are the relevant
ones, whereas those with  $k>k_B$ are neglected as these modes are
always  {\it inside} $\lambda$, i.e., $k^2<a''/a$ is never satisfied.
As we move away from the bounce, lower energy modes with $k<k_B$
reenter $\lambda$ and are also integrated.  To compute the energy of
produced particles, one can use Eq.~\eqref{rhov0}, but keeping in mind
that this expression is valid when we compare far past initial and far
future final states. Equation~\eqref{rhov0} then  gives the net energy
production between these states.  However, selecting the relevant
modes, $k_\mathrm{min}(\eta) \leq k\leq k_{B}$, we obtain
Eq.~\eqref{rhofinal}, which accounts only for particle modes.   In
Sec.~\ref{secIVA}, we use Eq.~\eqref{rhofinal} in order to obtain
an analytical estimate due to its simplicity.  However, for more
accurate results we need to observe that the initial condition is not
specified in the far past, but near the bounce. This may result in
some spurious energy density contribution in the initial vacuum, which
must be subtracted.  In this context, we can obtain an exact energy
density of produced particles using Eq.~\eqref{rhoEM}, which is
obtained directly from the energy-momentum tensor and is integrated
along the same lines of Eq.~\eqref{rhofinal}. However, this expression
does not  give the net result but the energy density at each instant
of time.  To obtain a net result, we must subtract the energy density
at the initial vacuum state.  {}Finally, we discuss the aforementioned
subtraction and the initial condition.  We take our initial condition
before the bounce, at the instant $t_i$ where $a''/a=0$.  This is not
too far from the bounce to guarantee the precision of
Eq.~\eqref{rhofinal}.  Also, the fact that  $a''/a=0$ at $t=t_i$ does
not guarantee that $a'/a=0$ in Eq.~\eqref{rhoEM}, which may result in
non-negligible contributions to the energy density.  However, if we
consider the subtraction of the vacuum energy for both expressions,
the computations are both consistent.  Either way, we have explicitly
verified that if the above cares are properly taken into account,
the results coming from either Eq.~\eqref{rhofinal} or
Eq.~\eqref{rhoEM} are completely equivalent. 

Considering the above remarks, the full set of coupled equations to be
solved numerically is then
\begin{eqnarray}
  \label{sys1}&& {\cal H}^2\equiv \left(\frac{a'}{a}\right)^2 =
  \frac{8\pi }{3 m_{\rm Pl}^2} a^2 \rho \left(1- \frac{\rho}{\rho_c}
  \right),\\
\label{sys2}&&\rho = \frac{\phi'^2}{2 a^2} + V(\phi) + \rho_p,\\
\label{sys3}&&\phi'' + 2 {\cal H}  \phi' + a^2\,{V}_{,\phi}=0,\\
\label{sys4}&&\chi_k'' + \left(k^2 - \frac{a''}{a}\right)\chi_k=0,\\
\label{sys5}&&\rho_p=\frac{1}{2\pi^2a^4}
\int_{\sqrt{\frac{a''(\eta)}{a(\eta)}}}^{k_B} dk\, k^2 \omega_k
|\beta_k(\eta)|^2,\\
\label{sys6}&& |\beta_k(\eta)|^2 = \frac{1}{2\omega_k}
\left[|\chi_k'(\eta)|^2 + \omega_k^2|\chi_k(\eta)|^2\right]
-\frac{1}{2}.
\end{eqnarray}

With the bounce dominated by the kinetic energy of the inflaton, the
potential $V(\phi)$ can be neglected during the contraction, bounce
and transition phases, and therefore, the explicit form of the
potential does not affect the particle production phase. In practice,
we consider the chaotic quartic potential for the
inflaton\footnote{The use a quartic inflaton potential here is only
  meant for illustrative purposes. Any other suitable inflaton
  potential could also be considered as well. The difference between
  others choices of inflaton potential is expected only to be mostly
  important on the slow-roll inflationary dynamics and on the
  resulting observable quantities. Even though the quartic inflaton
  potential is ruled out in the simple inflationary scenarios, it can
  lead to consistent results and be in accordance to the most recent
  Planck data in the context of the warm  inflation scenario (see,
  e.g., Refs.~\cite{Benetti:2019kgw,Graef:2018ulg} and references
  therein), where inflation happens in a state characterized by the
  presence of  dissipation and radiation effects.},
\begin{equation}
V(\phi)= \frac{V_0}{4} \left(\frac{\phi}{m_{\rm Pl}}\right)^4,
\label{pot}
\end{equation}
with $V_0$ fixed by the cosmic microwave background normalization for
the amplitude of the power spectrum~\cite{Aghanim:2018eyx}, which
gives $V_0/m_{\rm Pl}^4  \simeq 1.37 \times 10^{-13}$. 

The initial conditions for the $\chi$ field momentum modes are chosen
to be the BD ones,
\begin{eqnarray}\label{ICSchi}
\!\!\!\! \chi_k(\eta_i)=  \frac{1}{\sqrt{2k}}
e^{-ik\eta_i},\;\;\;\chi_k'(\eta_i)=  -i\sqrt{\frac{k}{2}}
e^{-ik\eta_i},
\end{eqnarray}
where the initial time $\eta_i$ is chosen in the prebounce phase,
when $a''/a=0$, which, in terms of the cosmic time, is found to be
$t_i \simeq -0.312 t_{\rm Pl}$. The initial conditions for the
inflaton field and its derivative are chosen such that we would have
around $60$-efolds of inflation when neglecting the backreaction of
the GPP.  This gives $\phi(t_i) \simeq 1.94 m_{\rm Pl}$ and
$\dot{\phi}(t_i) \simeq 0.45 m_{\rm Pl}^2$ for the case of the
potential in  Eq.~(\ref{pot}).

\begin{figure}[htb!]
\centerline{	\includegraphics[width=8.2cm]{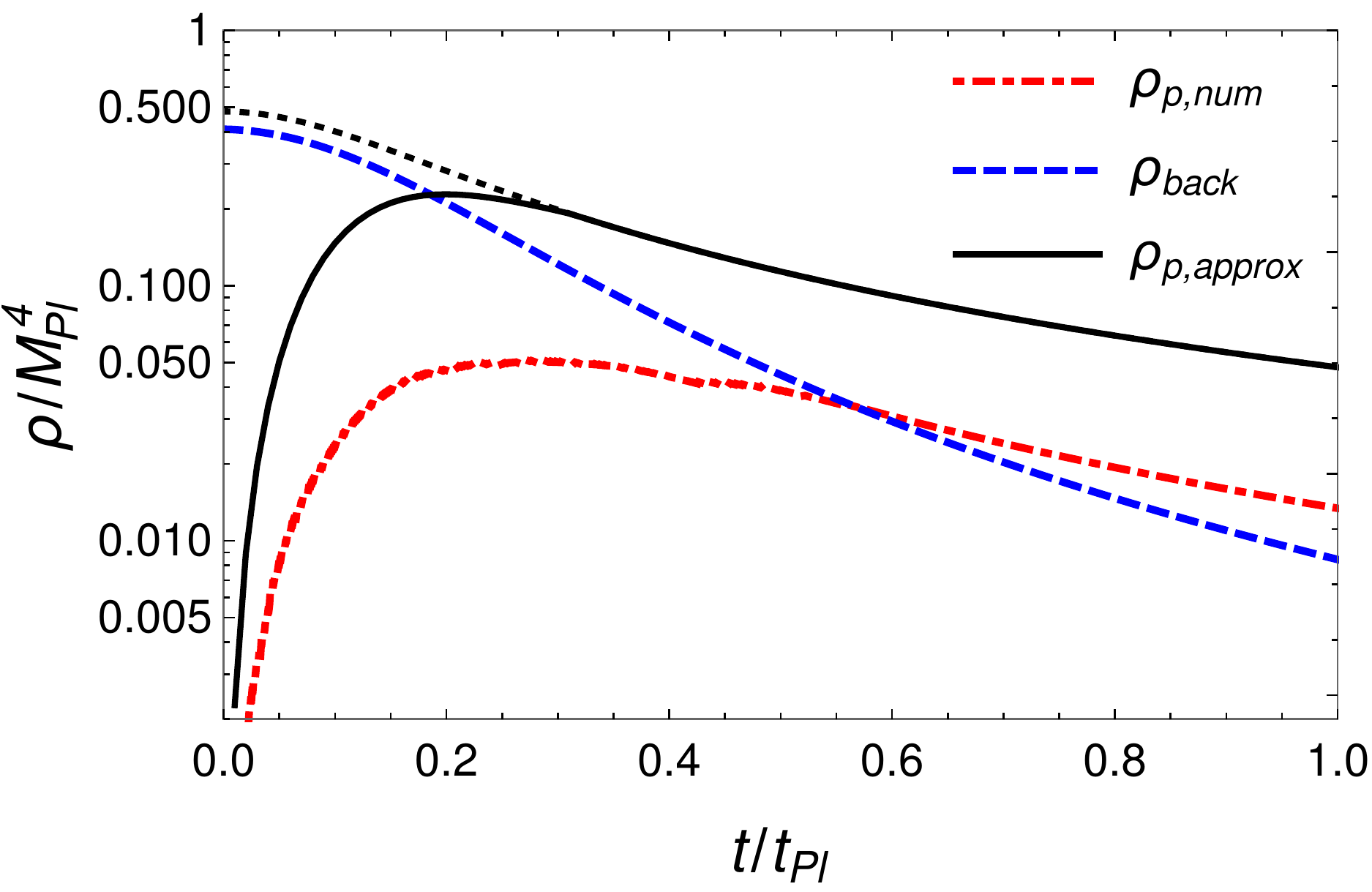} }
	\caption{Evolution of the different energy densities  as a
          function of the cosmic time.  The dot-dashed curve is the
          result obtained by numerically solving the coupled set of
          Eqs. (\ref{sys1})-(\ref{sys6}).  The dashed curve is
          the background energy density.  The solid curve is the
          result obtained when using the approximation given by
          Eq.~(\ref{intfromzero}),  while the dotted curve  is the
          result given by Eq.~(\ref{rhoGPP}).}  
	\label{densities}
\end{figure}

In {}Fig.~\ref{densities} we present the behavior of the energy
densities relevant for our analysis.  The dot-dashed curve is the
result for the energy density of produced particles by numerically
solving the coupled set of Eqs.~(\ref{sys1})-(\ref{sys6}).  The
dashed curve gives the background energy density.  The solid curve is
the result obtained when using the approximation given by
Eq.~(\ref{intfromzero}),  while the dotted curve is given by
Eq.~(\ref{rhoGPP}). We see that the simple result Eq.~(\ref{rhoGPP})
starts to agree with the one from Eq.~(\ref{intfromzero}) for
$t\gtrsim 0.3 t_{\rm Pl}$. In either case, both results overestimate
the one obtained from the complete numerical analysis, which
consistently accounts for the backreaction of the GPP. Both results,
however, show a peak around $t \sim 0.2 t_{\rm Pl}$, right after the
bounce and the superinflation phase.  The consistent inclusion of the
backreaction effect smoothly introduces the produced radiation in the
background, ceasing the continuous  particle production. In the
numerical calculations, $\rho_{p}$ should still grow for a moment
after the modes enter $\lambda$, due to the time interval for the
stabilization of the modes. By the time $t \gtrsim 0.6\, t_{\rm Pl}$,
all the modes are already stabilized, the GPP ceases and $\rho_{p}$
just redshifts away as relativistic matter, as expected. At its peak
value, we have $\rho_p \simeq 0.05 m_{\rm Pl}^4$, which is still
considerably larger than the upper bound estimated in the previous
subsection, $\rho_\chi \lesssim 2 \times 10^{-5}m_{\rm Pl}^{4}$.  By
extending the validity of  the dressed metric approach, the radiation
gravitationally produced~\footnote{Here, we denote as radiation all the
  contributions, which have equation of state $w=1/3$.} during the
bounce phase in LQC should not dominate the energy content of the
Universe before inflation starts. Nevertheless, the numerical
analysis, including the backreaction effects, shows that the radiation
domination phase will start already at around $t \sim 0.5\, t_{\rm
  Pl}$. 

\begin{center}
\begin{figure}[!htb]
\subfigure[]{\includegraphics[width=7.3cm]{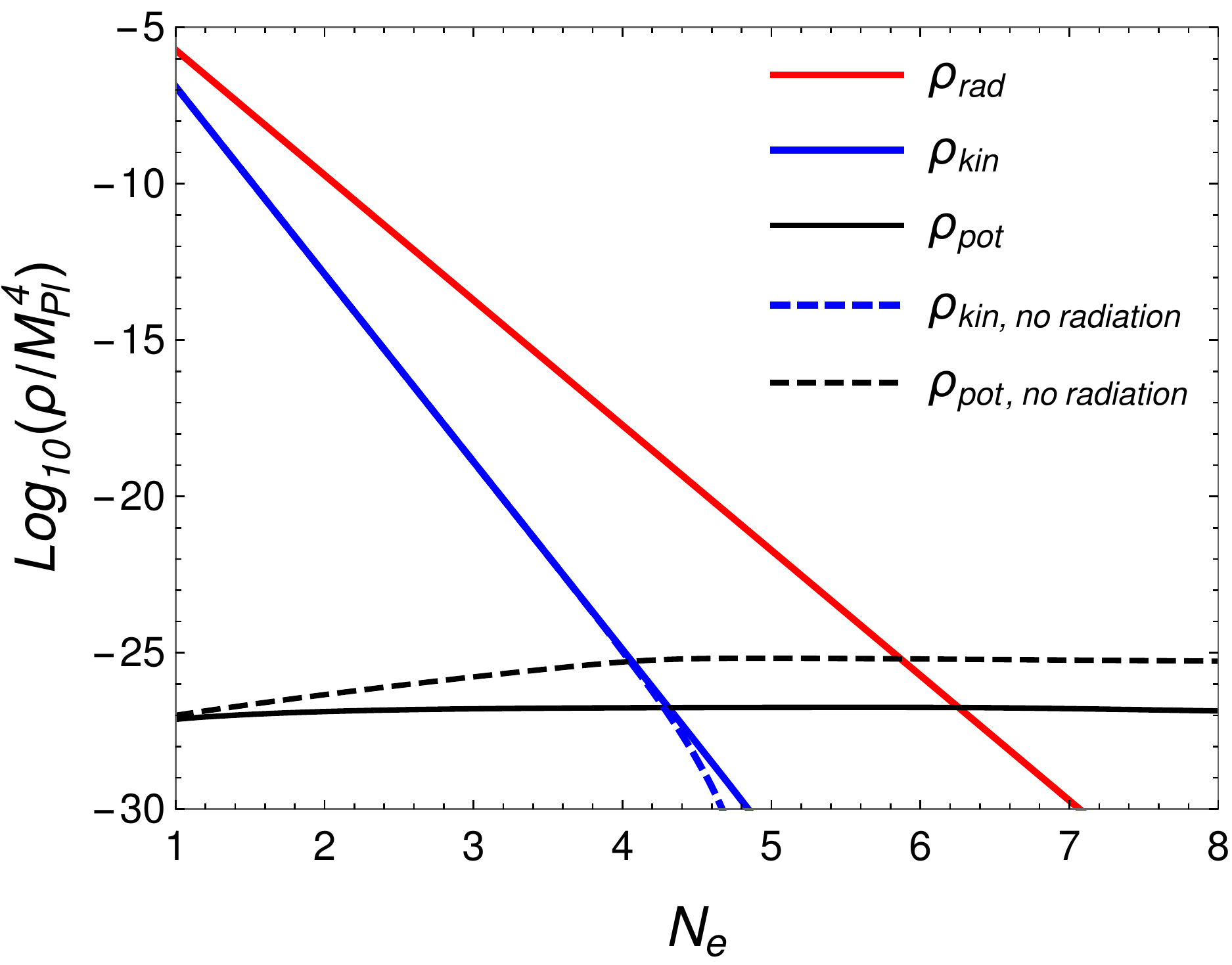}}
\subfigure[]{\includegraphics[width=7.3cm]{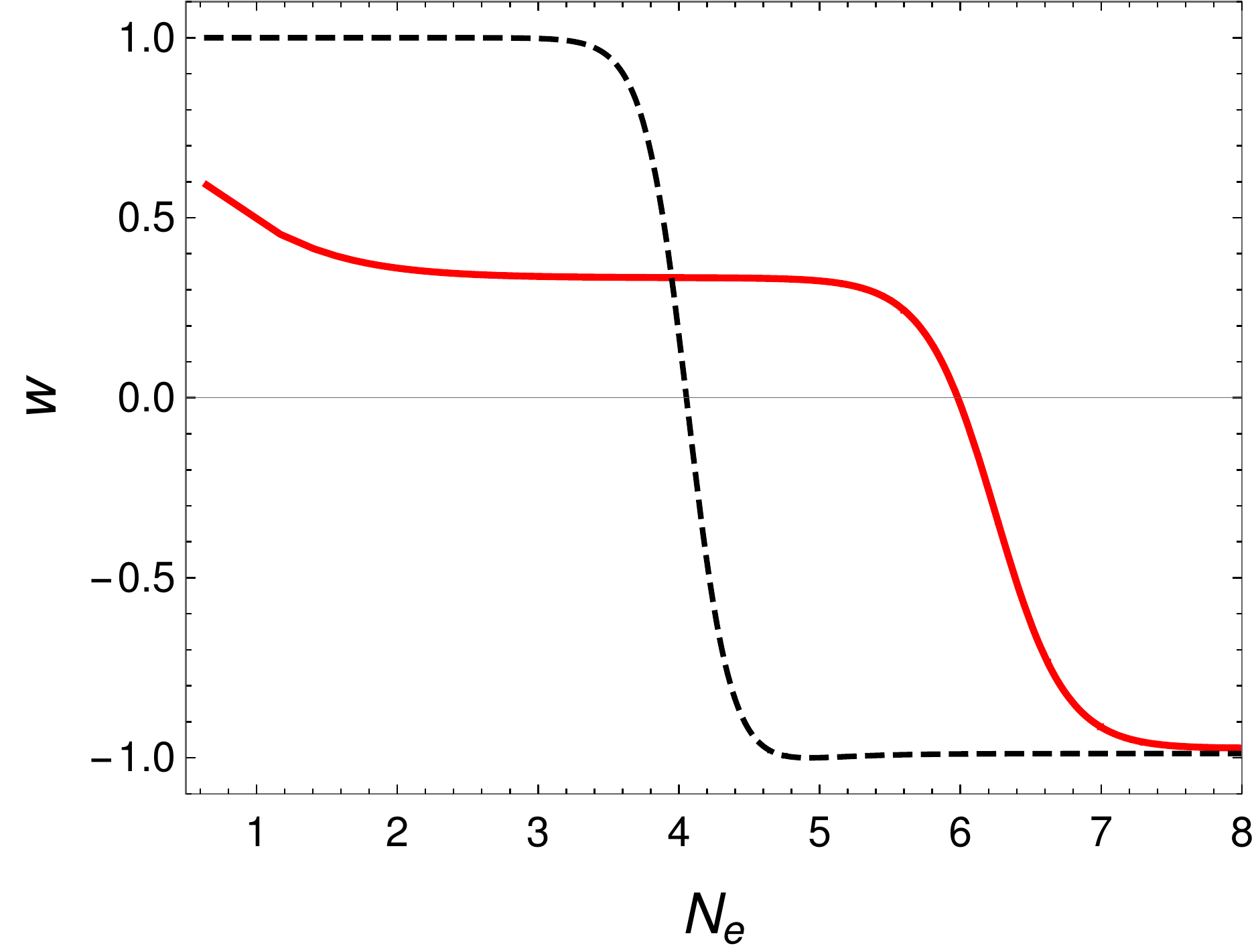}}
\caption{Panel (a): The radiation energy density due to GPP, the
  inflaton's kinetic energy $\dot \phi^2/2$ and potential energy
  $V(\phi)$  as a function of the number of \textit{e}-folds, in the presence of
  the backreation effect due to the GPP (solid lines), and in the
  absence of GPP (dashed lines). Panel (b): The equation of state
  $w=p/\rho$, in the presence of GPP (solid line), and in the absence
  of it (dashed line).}
\label{fig2}
\end{figure}
\end{center}

\begin{figure}[htb!]
\centerline{	\includegraphics[width=7.3cm]{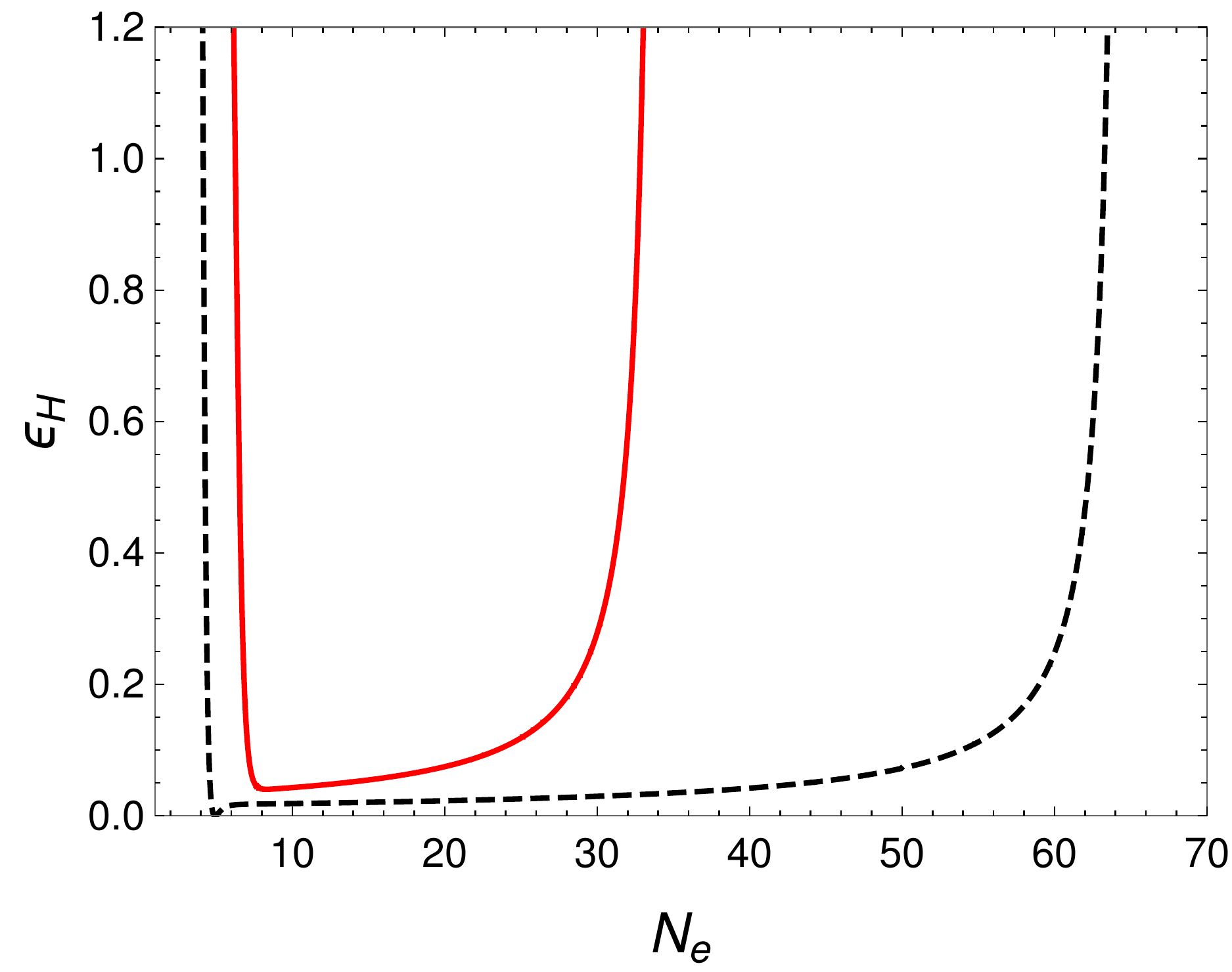} }
	\caption{The Hubble slow-roll parameter $\epsilon_H = - \dot
          H/H^2$, from the preinflationary phase up to the end of
          inflation, as a function of the number of \textit{e}-folds, in the
          presence of the  effect due to GPP (solid line) and in the
          absence of GPP (dashed line).}  
	\label{fig3}
\end{figure}

By extrapolating the evolution up until the start of the slow-roll
phase, in {}Fig.~\ref{fig2}(a), we show the evolution of the kinetic
and potential energies densities for the inflaton along with the
radiation energy density due to the GPP and that include all the
effects of backreaction due to the GPP in their respective evolutions
(solid lines). {}For comparison, we also show the results for the
kinetic and potential energies densities in the absence of the effects
of the GPP. In {}Fig.~\ref{fig2}(b) we show the equation of state
$w=p/\rho$, in the presence of GPP (solid line), and in the absence of
it (dashed line). The effect of the GPP is clear in the evolution of
the quantities shown in {}Fig.~\ref{fig2}. The energy density due to
GPP quickly dominates over the kinetic energy soon after the bounce as
described above. The kination regime in the preinflationary phase
after the bounce is soon replaced by  a radiation dominated regime ($w
\simeq 1/3$) that lasts until the potential energy dominates, when
the slow-roll phase starts.  While in the absence of GPP the
preinflationary phase lasts around four \textit{e}-folds, the GPP extends it to
around six \textit{e}-folds, thus delaying the start of the inflationary
phase. This also reflects strongly on the duration of the inflationary
phase, which can be seen in {}Fig.~\ref{fig3}, where we show the
slow-roll parameter $\epsilon_H = - \dot H/H^2$, already in the
preinflationary phase, till the end of inflation. In the absence of
GPP and for the choice of initial conditions we have taken in our
numerical example, we have 60 \textit{e}-folds of inflation. However, the GPP
shortens the inflationary phase to around 30 \textit{e}-folds.  This is a common
feature expected to happen to other choices of initial conditions for
the inflaton: there will be an increase of the duration of the
preinflationary phase and a decrease of the inflationary one, whenever
radiation dominates the dynamics after the
bounce~\cite{Graef:2018ulg}.

\section{Conclusions} 
\label{sec5}

We have investigated the process of gravitational particle production
in the preinflationary phase of loop quantum cosmology (LQC). We
show, for the first time, that if we suppose the validity of the
dressed metric approach in a LQC scenario with a BD initial condition in
the contracting phase, which is equivalent to the choice of a fourth 
order adiabatic vacuum at the bounce~\cite{Zhu:2017jew}, the
backreaction of the produced particles leads to a short radiation
dominated phase before the onset of inflation. Both the numerical and
analytical analysis agree qualitatively in this conclusion. In a
sense, the cosmological scenario we obtain in this case is very
similar to the one obtained for LQC in warm inflation, as shown e.g.,
in Refs.~\cite{Benetti:2019kgw,Graef:2018ulg}. In warm inflation, there
is also a preinflationary phase following the bounce that is
radiation dominated instead of kination dominated. However, in warm
inflation, the radiation is produced by the intrinsic dissipative
processes that can drive the decay of the inflaton field in light
(relativistic) degrees of freedom. Here, the radiation dominated
regime in the preinflationary phase after the bounce is due entirely
to gravitational particle production that occurs during the bounce
phase. 

In fact, our results give an indication that it is not consistent to
ignore the backreaction  of the gravitationally produced particles in
the preinflationary epoch of LQC, and this backreaction  leads to a
state at the onset of inflation  significantly different from the BD
vacuum. Since one should not  expected that the dressed metric
approach should be  self-consistent in this regime, our analysis put
in check the validity of the dressed metric approach in a LQC model
with such initial conditions. 

\section*{Acknowledgement}

The authors are thankful to A. Wang for discussions concerning the
topic of this work.  L.L.G acknowledge  the support from Conselho
Nacional de Desenvolvimento Cient\'{\i}fico e Tecnol\'ogico (CNPq),
Grant No.  307052/2019-2.  R.O.R. is partially supported by research
grants from CNPq, Grant No. 302545/2017-4, and Funda\c{c}\~ao Carlos
Chagas Filho de Amparo \`a Pesquisa do Estado do Rio de Janeiro
(FAPERJ), Grant No. E-26/202.892/2017. 


\end{document}